\documentclass[12pt,preprint]{aastex}
\usepackage{psfig}
 
\usepackage{psfig}

\def\deg      {{\ifmmode^\circ\else$^\circ$\fi}} 

\slugcomment{to appear in the Astrophysical Journal}
 
\shorttitle{Star formation rates in Lyman break galaxies.}
\shortauthors{Carilli et al.}
 
\begin{document}
  
 \title{Star formation rates in Lyman break galaxies: radio stacking of 
LBGs in the COSMOS field and the sub-$\mu$Jy radio source population.}
 
\author{ 
C.L. Carilli\altaffilmark{1},
Nicholas Lee\altaffilmark{1},
P. Capak\altaffilmark{2},
E. Schinnerer\altaffilmark{3},
K.-S. Lee\altaffilmark{4},
H. McCraken\altaffilmark{5},
M.S. Yun\altaffilmark{4},
N. Scoville\altaffilmark{2},
V. Smol{\v c}i{\' c}\altaffilmark{2},
M. Giavalisco\altaffilmark{4},
A. Datta\altaffilmark{1},
Y. Taniguchi\altaffilmark{6}
C. Megan Urry\altaffilmark{7,8}
}

\altaffiltext{$\star$}{Based on observations in the COSMOS Legacy Survey 
including those taken on the HST, Keck, NRAO-VLA, Subaru, KPNO 4m, CTIO 4m,
and CFHT 3.6m. The Very Large Array of the National Radio Astronomy
Observatory, is a facility of the National Science Foundation
operated under cooperative agreement by Associated Universities, Inc}

\altaffiltext{1}{National Radio Astronomy Observatory, P.O. Box 0, Socorro, NM
87801-0387}
\altaffiltext{2}{California Institute of Technology, MC 105-24, 1200 East
California Boulevard, Pasadena, CA 91125}
\altaffiltext{3}{Max Planck Institut f\"ur Astronomie, K\"onigstuhl
  17, Heidelberg, D-69117, Germany}
\altaffiltext{4}{Dept. of Astronomy, University of Massachusetts, Amherst, MA}
\altaffiltext{5}{IAP, Paris, France}
\altaffiltext{6}{Graduate School of Science and Engineering, Ehime University, 
Bonkyo-cho, Matsuyama, 790-8577m Japan}
\altaffiltext{7}{Department of Astronomy, Yale University, New Haven, CT 06511, USA}
\altaffiltext{8}{Yale Center for Astronomy and Astrophysics, Yale University, P.O.~Box 208121, New Haven, CT 06520}
 
\begin{abstract}

We present an analysis of the radio properties of large samples of
Lyman Break Galaxies (LBGs) at $z \sim 3$, 4, and 5 from the COSMOS
field.  The median stacking analysis yields a statistical detection of
the $z \sim 3$ LBGs (U-band drop-outs), with a 1.4 GHz flux density of
$0.90 \pm 0.21 \mu$Jy. The stacked emission is unresolved, with a size
$< 1"$, or a physical size $< 8$kpc.  The total star formation rate
implied by this radio luminosity is $31\pm 7$ $M_\odot$ year$^{-1}$,
based on the radio-FIR correlation in low redshift star forming
galaxies.  The star formation rate derived from a similar analysis of
the UV luminosities is 17 $M_\odot$ year$^{-1}$, without any
correction for UV dust attenuation.  The simplest conclusion is that
the dust attenuation factor is 1.8 at UV wavelengths. However, this
factor is considerably smaller than the standard attenuation factor
$\sim 5$, normally assumed for LBGs. We discuss potential reasons for
this discrepancy, including the possibility that the dust attenuation
factor at $z \ge 3$ is smaller than at lower redshifts.  Conversely,
the radio luminosity for a given star formation rate may be
systematically lower at very high redshift. Two possible causes for a
suppressed radio luminosity are: (i) increased inverse Compton cooling
of the relativistic electron population due to scattering off the
increasing CMB at high redshift, or (ii) cosmic ray diffusion from
systematically smaller galaxies.  The radio detections of individual
sources are consistent with a radio-loud AGN fraction of 0.3\%. One
source is identified as a very dusty, extreme starburst galaxy (a
'submm galaxy').

\end{abstract}
 
 \keywords{galaxies: formation --- galaxies: evolution --- galaxies:
radio --- surveys}
  
\section{Introduction}

The power of discovering high redshift galaxies via the broad-band
drop-out technique, ie. Lyman Break galaxies (LBGs), is now well
established (Steidel et al. 1999; 2000). Using this technique, over a
thousand galaxies have now been detected at $z > 2$.  Detailed studies
show that these galaxies have stellar masses between 10$^{10}$ and
10$^{11}$ $M_\odot$, and a (comoving) volume density at $z \sim 3$ of
$\sim 0.005$ Mpc$^{-3}$ (Giavalisco 2002).

A number of issues are still being investigated for LBGs. One
important issue is deriving the total star formation rate. Besides
model parameters such as the IMF and star formation history, the dust
correction in the UV remains under investigation.  Steidel et
al. (1999) originally estimated a typical UV dust attenuation factor
of about 5, based on optical spectroscopy.  Adelberger \& Steidel
(2000) applied a similar method to a larger sample of LBGs, as well as
observations at other wavebands (radio, submm), and conclude: "...the
mean extinction at 1600 \AA ~ for LBGs, a factor of 6 in our best
estimate, could lie between a factor 5 and a factor of 9."  More
recently, Reddy \& Steidel (2004) have derived the average dust
correction factors for UV-selected galaxies at $z \sim 2$ based on
deep X-ray, radio, and optical spectroscopic studies of galaxies in the
GOODS north field. They find that, for galaxies with total star
formation rates $> 20$ $M_\odot$ year$^{-1}$, the dust attenuation
factor is between 4.4 and 5.1.

A second issue for LBGs is the AGN fraction. Shapley et al. (2003)
found that $\sim 3\%$ of LBGs at z$\sim$3 have optical emission line
spectra that are consistent with AGN.

The Cosmic Evolution Survey (COSMOS), covering 2 \sq\deg, is a
comprehensive study of the evolution of galaxies, AGN, and dark matter
as a function of their cosmic environment. The COSMOS field has
state-of-the-art multiwavelength observations, ranging from the radio
through the X-ray (Scoville et al. 2007; Capak et al. 2007). A major
part of this study is to identify the largest samples of LBGs to date.
In section 2 we describe the LBG samples from the COSMOS survey based
on U, B, and V-band drop-out searches.

In this paper we consider the radio properties of these large samples
of LBGs, similar to the study of $z = 5.7$ Lyman-$\alpha$ emitting
galaxies by Carilli et al. (2007). Observations of the COSMOS field
have been done at 1.5$"$ resolution (FWHM) at 1.4 GHz with the Very
Large Array (Schinnerer et al.  2007). We employ the new VLA image
that has added integration time in the central $\sim 1^o$ (Schinnerer
in prep.), giving an rms at the field center of $\sim 7$ $\mu$Jy
beam$^{-1}$.  In section 3.1 we search for radio counterparts to
individual LBGs, while in section 3.2 we perform a stacking analysis
to derive the statistical properties of the samples.  In section 4 we
discuss the implications these observations have on the questions of
dust extinction and the AGN fraction in LBGs. We adopt a standard
concordance cosmology.
 
\section{A conservative COSMOS LBG sample}

We have identified large samples of U,B, and V-band drop-out galaxies
using the COSMOS photometric data (Capak et al. 2007).  For an
initial, robust analysis of the radio properties of high z LBGs, we
have adopted a series of conservative criteria to identify LBGs at the
different redshifts.  The U, B, and V-band drop selection criteria are
give in Table 1, where u' = CFHT u, i' = Subaru i, and r' = Subaru r
(Lee et al. in prep.).

It is well documents that LBG samples can include contaminants
(see Steidel et al. 2000, Capak et al. 2004, Bouwens et
al. 2007).  In the U and B-band drop-out selection, the contamination
comes from A and M stars along with $z<0.3$ star forming galaxies,
while the V-band drop-out selection is contaminated with M and T stars
along with $1.5<z<2.5$ obscured galaxies.  Since these contaminating
objects can be sources of radio emission, we take several steps to
produce a clean sample of high redshift objects.  We begin by removing
all ACS point sources with a star- or AGN-like spectral energy
distribution using the star/galaxy/AGN classifier described in Robin
et al. (2007).  In addition we remove all stars identified by a BzK
diagram (Daddi et al. 2007) at $K_{AB}<22$, this removes very red
stars which may not be identified by the Robin et al. (2007)
classifier.  We then use the BPZ photometric redshifts from Mobasher
\& Mazzei (2000) to remove objects with a best fit photo-z at $z<2.5$,
$z<3.5$, and $z<4.5$ for the U, B, and V-band drop-outs, respectively.
Finally we remove objects with a $>5$\% probability of being at
$z\leq1$.

The resulting sample is largely clear of contaminants.  Redshifts
were obtained with DEIMOS on Keck-II using the 830l/mm grating and
3.5-4h of integration time for a sample of color selected objects with
$i<26$ and $0.6<z<1.4$ or $z>4$ (Capak et al. in Prep).  The
spectroscopic sample matches 2 U-band drop-outs, 18 B-band drop-outs,
and 7 V-band drop-outs in our cleaned LBG sample.  Of these objects,
only 1 V-band drop-out object, which is a faint M star, is not in the
expected redshift range, yielding a contamination fraction of
$4\pm4$\%.  However, the U-band drop-outs sample was not specifically
targeted, so the constraint on the contamination fraction for these
objects is weak.  Of the remaining 26 objects at $z>3$, one B-band
drop-out shows an AGN signature, implying a residual AGN fraction of
$4\pm4$\%.

Although our sample is clean, it is not complete.  In general, the
phot-z cut creates a bias towards galaxies which have good photometry,
and closely match the expected UV-Optical spectral templates used for
the photometric redshifts.  Objects with unusual spectra, such as
AGN/galaxy composites, or objects with reddening that does not conform
to the Calzetti et al. (1997) or Milky Way laws, will likely be
excluded.  Furthermore, we have explicitly excluded optically bright
broad line AGN, and the probability requirements systematically bias
us against faint ($i<25$) objects.  The flux bias occurs because faint
objects have broad probability distributions due to their poor
photometric measurements.

We are also potentially biased against heavily obscured objects which
will have fainter UV fluxes, and hence poor phot-z fits.  An effort to
use intermediate band photometry (Ilbert et al. in prep) and a
systematic deep spectroscopic follow-up (Lilly et al. 2007, Capak et
al. in prep) is underway to fully characterize this sample and
understand the biases.

Table 2 summarizes the current samples. Column 2 indicates the
redshift range implied for a given drop-out band. Column 3 shows the
number of sources in the sample. Even with these conservative
selection criteria, we identify thousands of high redshift LBGs due to
the very wide COSMOS field.  In this initial study, we feel it
important to work with such a conservative sample of sources, in order
to make robust conclusions concerning the radio properties of LBGs.

For reference, the radio luminosity density at a rest frame frequency
of 1.4 GHz, for a source with $S_{\rm 1.4GHz} = 60\mu$Jy, assuming a
spectral index of $-0.8$ (Condon 1992), at $z \sim 3$ is: $L_{\rm
1.4GHz} = 3 \times 10^{31}$ ergs s$^{-1}$ Hz$^{-1}$.  For comparison,
the nearby radio AGN source M87 has $L_{\rm 1.4GHz} = 9\times 10^{31}$
erg s$^{-1}$ Hz$^{-1}$, while the luminous starburst galaxy Arp 220
has $L_{\rm 1.4GHz} = 4\times 10^{30}$ erg s$^{-1}$ Hz$^{-1}$.  In
M87, the radio emission is powered by a relativistic jet emanating
from the supermassive black hole with $M_{BH} \sim 2.5\times 10^9$
$M_\odot$ (Gebhardt et al. 2000), while in Arp 220 the radio emission
is driven by star formation with a total star formation rate $\sim
500$ $M_\odot$year$^{-1}$ (Carilli \& Yun 2000; Yun \& Carilli
2001). Even at the sensitivity of the COSMOS survey, we are limited to
detecting either radio AGN sources, or extreme starburst galaxies,
ie. galaxies with star formation rates $>$ a few thousand $M_\odot$
year$^{-1}$.

\section{The Radio properties of Lyman Break Galaxies}

\subsection{Individual sources}

We have searched for radio emission from each of the LBGs in the
COSMOS field.  At each position we determine the surface brightness,
and the rms noise in the region. The median rms in the LBG fields was
$\sim 15\mu$Jy beam$^{-1}$ for all three samples, implying a median
$4\sigma$ detection limit of 60$\mu$Jy beam$^{-1}$.  Our radio images
have a synthesized beam FWHM = 1.5'', hence we have searched for radio
sources brighter than 4$\sigma$, within 1$''$ radius of the LBG
optical position.  Column 4 in Table 2 lists the number of 4$\sigma$
radio counterparts in each sample.  Roughly, for each drop-out band,
we detect radio counterparts for $\sim 0.6\%$ of the UV sources.

We need to correct for the number of detections we expect by chance in
such a survey, either due to image noise, or faint sources unrelated
to the LBG.  We calculate this correction factor by performing
searches with identical criteria as for the LBG positions, but at
random positions selected to be $20''$ away from the coordinates of
the LBG samples.  This should give us a random sampling of the COSMOS
field, with noise statistics that match the LBG regions.  We find 19
4$\sigma$ random detections in the U-band drop-out fields, 4 for the
B-band drop-outs, and 3 for the V-band drop-outs.  These correspond to
0.29\%, 0.28\%, and 0.49\%, respectively.  We apply these corrections
in column 5 of Table 2.
 
Overall, we detect radio counterparts to $\sim 0.6\%$ of the UV
sources, with the expected number of detections at random being $\sim
0.3\%$. These results imply that only a small fraction, $\sim 0.3\%$,
of the LBGs, are radio sources brighter than $\sim 4\sigma$,
where, on average $4\sigma \sim 60\mu$Jy.

\subsection{A stacking analysis}

We investigate the statistical radio properties of the LBGs through a
stacking analysis. We adopt the median stacking method used by White
et al. (2007) in their stacking analysis based on the FIRST radio
survey.  White et al. perform detailed calculations that show that a
median stacking analysis is superior to a mean stacking, since it is
robust to small numbers of bright sources, and it does not require any
maximum allowed flux density cut-off prior to stacking.  Moreover, they
show that: '...in a limit where almost all of the values in the sample
are small compared to the noise, it is straight-forward to interpret
the median stack measurements as representative of the mean for the
population.' The COSMOS sample obeys this criterion.

Figure 1 shows the images derived by a median stacking of all the
sources in each drop-out band. These images are derived by extracting
sub-images centered around each LBG with a size of $18''\times
18''$. We then 'stack' these sub-images, aligned at the center
(on-source) position, and derive the median surface brightness at each
pixel position.  The off-source pixels serve as test positions for a
random stacking analysis. We find that the mean values for the off
positions equal zero, to within the noise, as expected for random
positions. We also derive the statistical noise of the analysis from
these off-positions.

Column 6 in Table 2 lists the peak surface brightnesses derived at the
stacked LBG position, plus the rms off-source noise, for each sample.
We obtain a clear detection of radio emission in the U-band drop-out
sample, with a value of $0.9\pm0.21\mu$Jy. 
The other bands show marginal detections, although at $<
3\sigma$ value, and hence we consider these upper limits.

Our results are consistent with the upper limit found by Ivison et
al. (2007), who performed a 1.4 GHz stacking analysis of 107 $z \sim
3$ LBGs in the extended Groth strip (the AEGIS survey), for which they
obtained a median value of $-0.6 \pm 2.3\mu$Jy.

The stacked source is unresolved, with a size $< 1"$. The implied
physical size is $< 8$kpc, hence the typical source must be galaxy
size, or smaller.

\section{Discussion}

\subsection{The AGN fraction and massive starbursts}

At the sensitivity limits of current deep radio surveys, such as
COSMOS, detection of individual sources at $z > 2$ is limited to
either extreme starbursts (star formation rates $> 1000$ $M_\odot$
year$^{-1}$), or radio jet sources with luminosities comparable to
M87.

After correcting for random detections, we obtain a detection rate of
$\sim 0.3\%$.  Shapley et al. (2003) find an AGN fraction in LBG
samples of $\sim 3\%$, based on optical spectroscopy.  If all of
our detected sources are radio AGN, then the implied radio loud
fraction is $\sim 10\%$. Interestingly, a value of 10\% is the
canonical value for radio loud AGN based on studies of nearby
galaxies (eg. Ivezic et al. 2002; Petric et al. 2007).  

We emphasize caution in interpreting our results on the radio loud
fraction of AGN at high redshift, for a number reasons. First, in a recent
comparison of the FIRST and SDSS surveys, Jiang et al. (2007) find
that the radio loud fraction likely depends on both optical luminosity
and redshift. This study is not directly comparable to our COSMOS
study, since they were limited to much more luminous sources at high
redshift, by about two orders of magnitude in the radio.

Second, the above analysis does not consider the possibility that
some of the 0.3\% of the radio detections are extreme, highly obscured
starbursts, comparable to the submm galaxies. Indeed, in a companion
paper, we present the discovery of a submm galaxy in our
radio-selected LBG COSMOS sample (Capak et al. 2008). The
radio-detected LBG has a spectroscopic redshift of $z =4.5$, and a
radio flux density of $45\pm 10\mu$Jy. Thermal emission from warm dust
is detected from this galaxy at 250GHz using the  MAMBO
bolometer camera, with a flux density of $3.4\pm 0.7$mJy.  This
source is the first submm galaxy yet identified (spectroscopically) at
$z > 4$ (Capak et al. 2008).

And third are the caveats mentioned in section 2 concerning our
conservative source selection criteria, eg. excluding obvious
optically bright broad line AGN.

\subsection{The sub-$\mu$Jy radio source population}

The unprecedented size of the COSMOS LBG sample has allowed us to
reach sub-$\mu$Jy sensitivity levels in the stacking analysis of the
LBG samples. We obtain a clear statistical detection of $S_{\rm 1.4GHz} =
0.90\pm 0.21\mu$Jy for the U-band ($z \sim 3$) drop-outs. The implied
rest frame 1.4 GHz luminosity density is $L_{\rm 1.4\rm GHz} = 5.1\times
10^{29}$ erg s$^{-1}$ Hz$^{-1}$, assuming a spectral index of $-0.8$
(Condon 1992).

Assuming the radio emission is driven by star formation, we can derive
a total star formation rate (0.1 to 100 $M_\odot$), from the rest
frame 1.4 GHz luminosity density. There are a number of recent
calibrations of this relationship for nearby galaxies using, eg.  the
IRAS and NVSS surveys.  These studies adopt the star formation rates
derived from the far-IR emission based on the relations in Kennicutt
(1998), assuming a Salpeter IMF, and then calibrate the radio
conversion factor using the tight radio to far-IR correlation for star
forming galaxies.  We adopt the conversion factor from the study of
Yun, Reddy, and Condon (2001): 

$$\rm SFR = 5.9 \pm 1.8 \times 10^{-29} {\sl L}_{\rm 1.4GHz} ~
M_\odot~ year^{-1}, $$

\noindent where $L_{\rm 1.4\rm GHz}$ is in erg s$^{-1}$ Hz$^{-1}$.
From this, we calculate a mean star formation rate of $31\pm 7$
$M_\odot$ year$^{-1}$. Note that the conversion factor above is within
10\% of that derived by Bell (2003).

The mean observed UV luminosity at a rest frame wavelength of 1600 \AA
~of the U-band drop-out sample is: $L_{2000A} = 1.2\times 10^{29}$ erg
s$^{-1}$ Hz$^{-1}$ (Capak et al. 2008 in prep).  Using equation (1) in
Kennicutt (1998) for the relationship between UV luminosity and total
star formation rate, we derive a total star formation rate of 17
$M_\odot$ year$^{-1}$, uncorrected for dust attenuation.

The ratio of radio derived star formation rate to UV derived star
formation rate is $1.8\pm 0.4$.  The simplest conclusion is that the
UV emission is attenuated by dust by a factor of 1.8. However, this
factor is considerably smaller than the standard factor $\sim 5$
adopted for LBGs (see Section 1). We consider some possible reasons
for this difference.

One possibility is that the dust attenuation factor for the $z \sim 3$
COSMOS LBG sample is indeed smaller than for other LBG samples. Most
of the studies of dust attenuation of LBG galaxies have been at $z <
3$ (see section 1), and hence it is possible that at higher redshift
the dust attenuation factor decreases. An argument against this
decrease is the recent X-ray and optical study of a sample of LBGs at
$z \sim 3$ by Nandra et al. (2002), who also derive a dust attenuation
factor of about 5. We also re-emphasize that the COSMOS LBG sample
analyzed herein de-selected sources with poor phot-z fits, which can
occur for very heavily obscured objects (Section 2).

On the other hand, Wilkins, Trentham, \& Hopkins (2008) present a
detailed comparison of the build up of stellar mass in galaxies versus
the cosmic star formation rate density. They conclude that there is a
discrepancy between the rate of stellar mass creation, and the star
formation rates derived from the UV luminosities, at high redshift, if
one assumes the standard factor $\sim 5$ dust attenuation. The
discrepancy is in the sense that the UV derived star formation rates
are too high. They find that the peak in the discrepancy occurs at $z
\sim 3$, and they suggest that the UV-derived star formation rates at
this redshift may be over-estimated by a factor $\sim 4$. They also
point out that '...this large deviation at high redshift offers an
explanation for why the integrated star formation history implies a
local stellar mass density in excess of that measured.'

A second possibility is that the conversion factor of radio luminosity
to star formation rate is higher in the $z \sim 3$ LBG COSMOS sample
than has been derived for low redshift galaxies.  Radio studies of
24$\mu$m selected galaxies by Appleton et al. (2004) imply that the
radio conversion factor is constant out to $z \sim 1$, ie. that the
radio-FIR correlation is constant out to this redshift.  Reddy \&
Steidel (2004) extend this conclusion out to $z \sim 2$ in their
extensive study of UV selected galaxies. Most recently, Ibar et
al. (2008) conclude that the radio conversion factor remains constant
out to $z \sim 3$, using a 24$\mu$m source sample with redshifts from
the SXDF. However, given the need for individual source detections in
the radio, their $z \sim 3$ radio sources have star formation rates
about two orders of magnitude larger than the stacking results
presented herein, and hence a direct comparison is problematic.

One physical reason why we might expect the radio conversion factor to
diverge at the highest redshifts is increased relativistic electron
cooling due to inverse Compton scattering off the cosmic microwave
background (CMB). The ratio of relativistic electron energy losses due
to synchrotron radiation, to energy losses due to inverse Compton
radiation, equals the ratio of the energy density in the magnetic
field to that in the photon field.  The energy density in the CMB
increases as: $U_{CMB} = 4.2\times 10^{-13} (1+z)^4$ ergs cm$^{-3}$.
Figure 2 shows a comparison of $U_{CMB}$ with the typical energy
densities in the magnetic fields in different regions in galaxies. We
show the range of fields considered typical for spiral arms ($\sim$
few $\mu$G), and for starburst galaxy nuclei (of order 100$\mu$G; Beck
et al. 1994; see review by Beck 2005).  The important point is that IC
losses off the CMB will dominate synchrotron losses in a typical ISM
at $z \ge 0.5$, and dominate in starburst nuclei at $z \ge 4$.

We note that inverse Compton losses will not affect the thermal
electrons responsible for Free-Free emission from star forming
galaxies. Such Free-Free emission may dominate the total radio
emission from galaxies at rest frequencies between roughly 40 GHz and
100 GHz (Condon 1992), and hence become an important factor in radio
continuum studies of very high redshift galaxies.

A depressed radio luminosity for a given star formation rate could
also arise if the $z \sim 3$ LBGs are systematically smaller
galaxies. The standard model that produces the radio-FIR correlation
(Condon 1992) requires a cosmic ray processing box size $\ge 1$
kpc. It has been observed that dwarf galaxies at low redshift depart
from the radio-FIR correlation by about a factor of 2, in the sense of
being radio under-luminous.  The hypothesis is that the cosmic rays
diffuse out of the galaxy on timescales shorter than required to
maintain the standard radio-FIR correlation (Yun et
al. 2001). Giavalisco (2002) describes high $z$ LBGs as having typical
half-light radii of 4 to 7 kpc, significantly larger than typical
dwarf galaxies. However, he points out that: "..frequently the
galaxies have disturbed or fragmented morphologies, with one bright
core, or multiple knots embedded in diffuse nebulosity, reminiscent of
merger events."

In this paper, we report the first robust statistical (median)
detection of sub-$\mu$Jy radio emission from LBG galaxies at $z \sim
3$. This detection was made possible by the very wide area, and depth,
of the Cosmos field. While our physical interpretation of the result
remains inconclusive, there are a number of future studies we are
pursuing to address the interesting questions raised concerning the UV
attenuation factor for LBGs, and the the radio luminosity to star
formation rate conversion factor, at $z \ge 3$. The most important
study involves obtaining optical spectra of a large sample of LBGs
from the COSMOS sample. Spectra will elucidate the nature of the
sources, and allow for a study of the dust correction factor as a
function of galaxy type (AGN, starburst, elliptical...). Also, the
selection criteria for this LBG sample were very conservative (Section
2).  We will further refine (and increase) our LBG samples using the
Cosmos multiband photometry, and perform statistical analyses of the
radio properties as a function of eg. stellar mass, or $A_V$. We are
also exploring the IR properties of these samples with Spitzer. Such
IR studies have particular relevance in the light of the results of
Ivison et al. (2007), who find that for the seven IR luminous $z \sim
3$ LBGs in their AEGIS sample ($S_{24\mu m} > 60\mu$Jy), the median
1.4 GHz flux density is 44$\mu$Jy, comparable to that expected for
sub-mm galaxies. Lastly, unambiguous identification of radio AGN using
eg. optical spectra, will provide input into faint radio source
population models that are being generated in the context of planning
for future, large area radio telescopes (Wilman et al. 2008). The
results presented herein indicate that future deep (sub-$\mu$Jy), wide
field surveys with the Expanded Very Large Array, will detect
routinely the radio emission from individual LBGs out to $z \sim 3$.
 
\acknowledgments
 
The HST COSMOS Treasury program was supported through NASA grant
HST-GO-09822.  We gratefully acknowledge the contributions of the
entire COSMOS collaboration consisting of more than 70 scientists.
More information on the COSMOS survey is available \\ at {\bf
\url{http://www.astro.caltech.edu/cosmos}}.  CC thanks the
Max-Planck-Gesellschaft and the Humboldt-Stiftung for support through
the Max-Planck-Forschungspreis. We thank the referee for helpful 
comments. 
 

\clearpage
\newpage

\begin{table}\label{LBGtable}
\begin{center}
\caption{COSMOS drop-out selection}
\begin{tabular}[ht]{|c|c|c|c|}
\tableline
U-band drop-outs & ($\mbox{u'} - B) \geq 0.7  (B - \mbox{i'})
+ 0.7$ & ($\mbox{u'} - B) \geq 1.4$ &  ($B - \mbox{i'}) \leq 2.5$ \\
B-band drop-outs & ($B-V) \geq 0.7~(V - \mbox{i'}) + 0.7$ &  ($B-V)
\geq 1.0$ & $(V - \mbox{i'}) \leq 2.5$ \\
V-band drop-outs & $(V-\mbox{r'}) \geq 0.3~(\mbox{r'}-\mbox{i'}) +1.0$ &  ($V-\mbox{r'})\geq 1.4$ & $(\mbox{r'}-\mbox{i'}) \leq 3.0 $\\
\tableline
\end{tabular}
\end{center}
\end{table}

\begin{table}\label{LBGtable}
\begin{center}
\caption{Statistics of Lyman Break Sample and Results of the $4\sigma$ Counterpart Search}
\begin{tabular}[ht]{|c|c|c|c|c|c|}
\tableline
Band & Redshift & \# of LBGs &  \# of $4\sigma$ Sources & 
Corrected \# of Sources & Median flux density \\ 
~ & ~ & ~ & ~ & ~ & $\mu$Jy \\
\tableline
U & 2.5--3.5 & 6457 &  43 (0.67\%)  & 24 (0.38\%) & $0.90\pm 0.21$ \\
B & 3.5--4.5 & 1447 &  8 (0.55\%)  & 5 (0.34\%) & $0.83\pm 0.42$ \\
V & $>$ 4.5 & 614 &    5 (0.81\%) & 2 (0.33\%) & $1.72\pm 0.68$ \\
\tableline
\end{tabular}
\end{center}
\end{table}

\begin{figure}
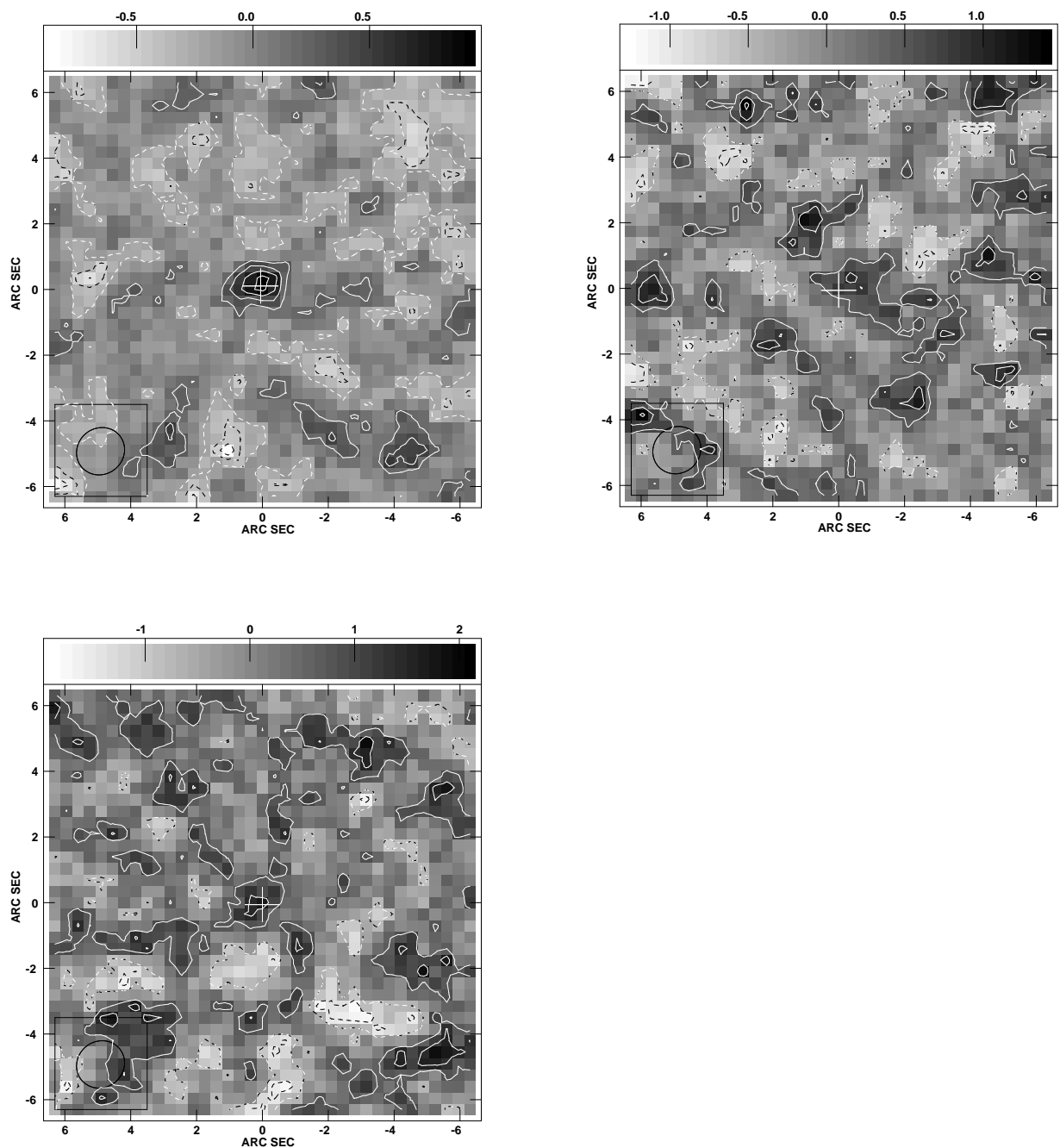

\psfig{file=f1a.ps,width=3in}
\vskip -3.74in
\hspace*{3in}
\psfig{file=f1b.ps,width=3in}
\psfig{file=f1c.ps,width=3in}
\caption{Images at 1.4 GHz of the median stacking results 
for the different drop-out samples. In each case, the contour
levels are -3, -2, -1, 1, 1, 2, 3, 4 $\times$ the rms value in the image, and
the cross indicates the source stacking position. The synthesized  
beam FWHM is shown in the lower left corner. 
Top left: U-band drop-out result, with rms = 0.21 $\mu$Jy beam$^{-1}$.
Top Right: The B-band drop-outs result, with rms = 0.42 $\mu$Jy beam$^{-1}$.
Bottom: The V-band drop-outs result, with rms = 0.68 $\mu$Jy beam$^{-1}$.
}
\end{figure}

\begin{figure}
\psfig{file=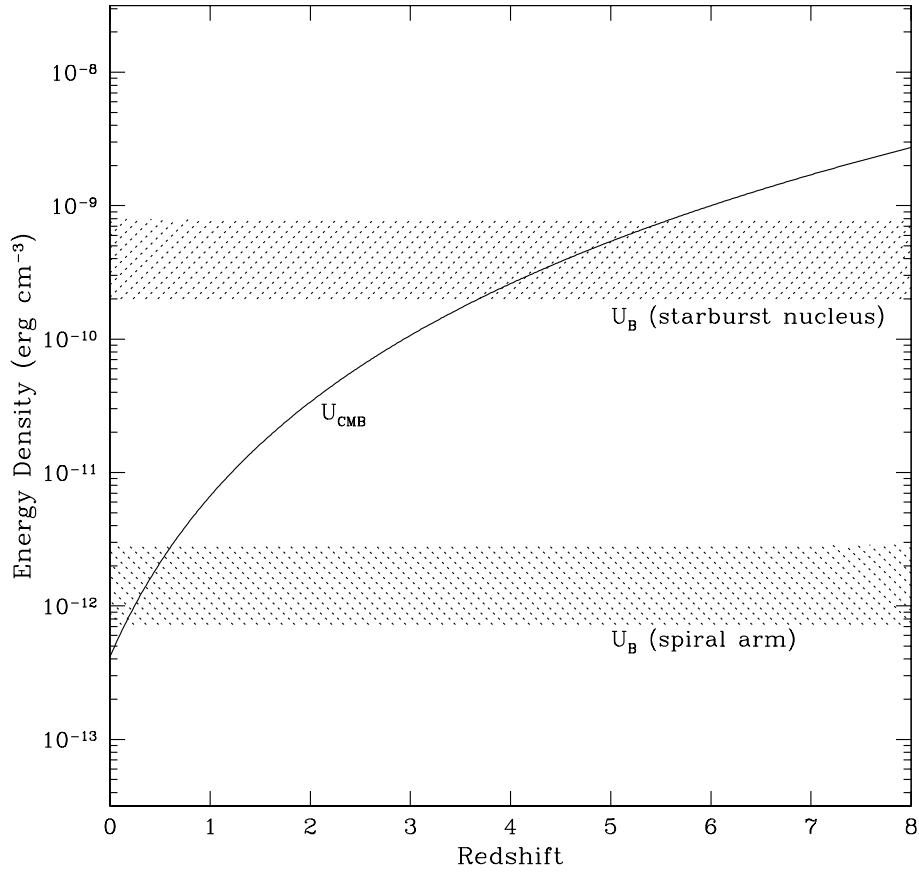,width=5in}
\caption{The energy density in the CMB as a function of
redshift, plus shaded regions showing typical ranges for magnetic
fields in spiral arms, and starburst galaxy nuclei.}
\end{figure}

\end{document}